\documentclass[aps,prb,twocolumn,lengthcheck,superscriptaddress,showpacs,preprintnumbers,amsmath,amssymb]{revtex4}

\usepackage{graphicx}
\usepackage{dcolumn}
\usepackage{bm}

\usepackage[T1]{fontenc}
\usepackage{ae,aecompl}

\begin{document}

\title{Strong-coupling expansions for the topologically inhomogeneous Bose-Hubbard model}

\author{P. Buonsante}
\affiliation{Dipartimento di Fisica, Politecnico di Torino and I.N.F.M, Corso Duca degli Abruzzi 24 - I-10129 Torino (ITALIA)}%
\author{V. Penna}%
\affiliation{Dipartimento di Fisica, Politecnico di Torino and I.N.F.M, Corso Duca degli Abruzzi 24 - I-10129 Torino (ITALIA)}%
\author{A. Vezzani}
\affiliation{Dipartimento di Fisica, Universit\`a degli Studi di Parma and I.N.F.M., Parco Area delle Scienze 7/a I-43100 Parma (ITALIA)}%

\date{\today}

\begin{abstract}
We consider a Bose-Hubbard model with arbitrary hopping term and provide the boundary of the  insulating phase thereof in terms of third-order strong coupling perturbative expansions for the ground state energy. In the general case two previously unreported terms occur, arising from triangular loops and hopping inhomogeneities, respectively. Quite interestingly the latter involves the entire spectrum of the hopping matrix rather than its maximal eigenpair, like the remaining perturbative terms. We also show that hopping inhomogeneities produce a first order correction in the local density of bosons. 
Our results apply to ultracold bosons trapped in confining potentials with arbitrary topology, including the realistic case of optical superlattices with uneven hopping amplitudes. Significant examples are provided. Furthermore, our results  can be extented to magnetically tuned transitions in Josephson junction arrays. 

\end{abstract}


\maketitle

\section{Introduction}
The Bose-Hubbard (BH) model describes a gas of interacting bosons hopping across the sites of a discrete structure. It was originally introduced\cite{A:Fisher} to study the the superfluid-insulator quantum phase transition in liquid helium, and later argued to describe the physics of an array of coupled Bose-Einstein condensates (BECs).\cite{A:Jaksch} 
Such proposal has been strikingly substantiated by a recent breakthrough experiment,\cite{A:Greiner} where  the superfluid-insulator transition was observed in a $^{87}$Rb BEC array.

This result greatly reinvigorated the studies on the quantum phase transition characterizing the BH model.\cite{A:Kashurnikov02,A:Batrouni02,A:Rey,A:Pupillo03,CM:Jain} Indeed, the ongoing progress in condensate trapping techniques promises an unprecedented control on the parameters of the experimental realizations of such model. A simple example in this respect is supplied  by the quite direct relation between the intensity of the laser beam trapping the condensates at the sites of the so-called {\it optical lattice} and the tunneling interaction between neighbouring condensates, accounted by the hopping amplitude in the relevant BH Hamiltonian.\cite{A:Jaksch,A:Greiner} 

Recently, several proposals have been brought forward concerning the realization of BEC arrays characterized by  non-trivial layouts. Suitable laser setups, possibly with different wavelengths, can be used to generate optical superlattices.\cite{A:Guidoni97,A:Roth03,A:Peil,A:Blakie,A:Santos} Techniques borrowed from holography\cite{A:Grier} or from semiconductor technology\cite{A:Folman01} have been proposed for the creation of optically or magnetically confined inhomogeneous BEC arrays.

The BH model is also commonly used to describe Josephson junction arrays\cite{A:FazioPR} (JJAs), which can be easily arranged according to non-trivial layouts, including inhomogeneous ones.\cite{A:Meyer02,A:Cataudella}
In this case the phase transition can be observed measuring different samples or adjusting an external magnetic field applied to the same sample.\cite{A:FazioPR}

In the light of this, the need arises for the study of a BH model characterized by a generic hopping term. A further point of interest in this respect is provided by the unexpected features induced by topological inhomogeneities in the properties of discrete boson systems\cite{A:BHComb} even in the absence of boson interaction.\cite{A:CombL,A:MesoComb}

For the study of the zero-temperature phase diagram of the BH model several techniques have been adopted. Qualitatively correct results are provided by numerical\cite{A:Sheshadri,CM:Jain} and analytical\cite{A:Amico,A:vanOosten,A:LobiMF} mean-field approaches. As it is well known, even for relatively small systems the formidable size of the Hilbert space rules out the route of direct diagonalization in the study of the ground state properties of the BH Hamiltonian, determining the features of the zero-temperature phase diagram. In the case of homogeneous systems, clever numerical methods have been devised that provide quite satisfactory results. However, since they rely on features such as 
self-similarity\cite{A:Kuehner} or  translational invariance, \cite{A:Elstner99a} their generalization to topologically inhomogeneous structures is at least problematic.
Of course, the superfluid-insulator transition on such structures can be studied  exploiting the flexibility of  Quantum Monte Carlo simulations, that provide the most general and reliable numerical approach. Indeed, this technique was successfully applied to the case of regular lattices\cite{A:Batrouni,A:Kashurnikov96}, possibly in the presence of inhomogeneities arising from random\cite{A:Batrouni92,A:Batrouni93,A:Kisker97,A:Lee01} or harmonic\cite{A:Kashurnikov02,A:Batrouni02} local potentials. Recently, some results addressing inhomogeneous hopping have been reported.\cite{A:BHComb}
As to analytical approaches, the strong coupling perturbative results reported in Ref.~\onlinecite{A:Freericks2} allow a quite satisfactory description of the zero-temperature phase diagram of the BH model on regular bipartite lattices, possibly with disorder.

Here, we extend the third-order perturbative results of Ref.~\onlinecite{A:Freericks2} to a generic structure,
 possibly non-bipartite and/or topologically inhomogeneous. We remark that commonly investigated regular lattices, such as the triangular\cite{A:Niemeyer,A:Blakie} or the Kagom\`e\cite{A:Santos} lattice, are non-bipartite.
 We show that in the general case  two further  significant third-order contributions appear in the perturbative expansion of the ground state energy of the system, and of  the boundaries of the Mott lobes thereof. Furthermore we show that terms arising from inhomogeneous hopping amplitudes appear in the expansion of a physically interesting quantity such as the local density of bosons already at the first perturbative order. We also supply some examples where the new terms play a significant role.

\section{The Bose-Hubbard model on a generic structure}
The Hamiltonian for the {\it pure} BH model on a generic structure comprising $M$ sites has the form \begin{equation}
\label{E:BH}
H =  \sum_{j=1}^M \left[\frac{U}{2} n_j (n_j-1)-\tau \sum_{h=1}^M a_j  t_{j h} a_h^+\right]
\end{equation}
where $a_j$ ($a_j^+$) is the boson annihilation (creation) operator at site $j$, $n_j=a_j^+ a_j$ is the relevant number operator, $U>0$ is the boson repulsion strength  and $\tau>0$ is an overall scaling factor controlling the magnitude of the possibly inhomogeneous hopping matrix, $t$. Since the latter contains information about the connectivity of the underlying structure, it is $t_{j h} = 0 \Leftrightarrow  A_{j h} = 0$, where $A$ is the so-called adjacency matrix, whose generic entry $A_{j h}$ equals one only  if sites $j$ and $h$ are adjacent, and zero otherwise. Hence topological inhomogeneity can come in the form of a non-trivial connectivity, i.e. a site dependent coordination number $z_j=\sum_h A_{j h}$, and/or in the form of uneven hopping amplitudes $t_{ij}$. 
We remark that this applies also to periodic structures, such as the inhomogeneous Kagom\`e lattice in Fig.~\ref{F:z4} (C.) or the $T_3$ lattice of Ref.~\onlinecite{A:Cataudella}. 
Note that Hamiltonian (\ref{E:BH}) commutes with the total number of bosons, $N=\sum_{j=1}^M n_j$. Hence the properties of its ground state $|\psi\rangle$ --- i.e. the state minimizing $\langle\psi |H-\mu N|\psi \rangle$, where $\mu$ is the chemical potential controlling the boson population  ---  can be conveniently studied considering one number eigenspace at a time.

The competition between the on-site repulsion and the kinetic energy --- proportional to $U$ and $\tau$, respectively --- 
produces a well-known zero-temperature phase diagram  in the $\mu/U$-$\tau/U$ plane, where a series of adjacent Mott-insulator lobes appears. Outside these lobes the cost of changing the total number of particles in the system by a finite amount vanishes in the thermodynamic limit.
Conversely, within the Mott lobes, adding or subtracting a single particle always costs a finite amount of energy, and the total boson population is commensurate to the size of the structure,  $N = n_0 M$, where $n_0 \in {\mathbb N}$.
Hence, the left and right boundary of the filling-$n_0$ Mott lobe are customarily evaluated as the ground-state energy gaps 
\begin{equation}
\label{E:mu}
\begin{array}{l}
\mu_{\rm L}(n_0,\tau) = E_{\rm Mott}(n_0,\tau)-E_{\rm hole}(n_0,\tau)\\
\mu_{\rm R}(n_0,\tau) = E_{\rm part}(n_0,\tau)-E_{\rm Mott}(n_0,\tau)
\end{array}
\end{equation}
where $E_{\rm Mott}(n_0,\tau)$, $E_{\rm hole}(n_0,\tau)$ and $E_{\rm part}(n_0,\tau)$ are the ground state energies of $H$ when the total population is $n_0 M$, $n_0 M -1$ and $n_0 M +1$, respectively. 
Following Ref.~\onlinecite{A:Freericks2}, in the sequel we term the latter two  {\it defect states} and label ``def'' quantities relevant to either of them.

In Ref.~\onlinecite{A:Freericks2} the quantities in Eqs.~(\ref{E:mu})
are quite satisfactorily approximated by a third-order strong coupling perturbative expansion, the perturbative parameter being $\tau$. Such results carry an explicit dependence on the hopping matrix $t$ and, quite interestingly, on its maximal eigenvalue and eigenvector. This means that the influence of topology, algebraically described by $t$, can be in principle  singled out explicitly in the perturbative expansion of $\mu_{\rm L}$ and $\mu_{\rm R}$.
However, the perturbative results  
reported in  Ref.~\onlinecite{A:Freericks2} apply only to homogeneous bipartite lattices, whose hopping matrix is such that $\sum_{h=1}^M t_{j h}=z$ and $(t^{2 p+1})_{h h}=0$. Actually, since such expansions are limited 
to the third order, the latter constraint can be weakened, and only regular lattices featuring triangular loops have to be excluded, $(t^3)_{h h}=0$. 
In this respect we mention that third order results have been reported for the particular case of the regular triangular lattice in the presence of a magnetic field.\cite{A:Niemeyer}

\section{Perturbative Results}
In the following we provide the complete third-order strong coupling perturbative expansions for the quantities in Eqs.~(\ref{E:mu}),
obtained including those terms arising from the presence of triangular loops and topological inhomogeneities in the discrete structure underlying the BH model. Such terms, pertaining to the third perturbative order, are denoted ``$\triangleleft$'' and ``$\rtimes$'', respectively. Quite interestingly, the latter depends not only on the maximal eigenpair  of the hopping matrix, but on its entire spectrum $t\, {\mathbf f}_k = \lambda_k {\mathbf f}_k $, $\lambda_1=\max_k(\lambda_k)$.

For commensurate fillings, $N= n_0 M$, the energy of the ground state is
\begin{equation}
\label{E:Mott}
E_{\rm Mott} = E^\diamond_{\rm Mott}+\tau^3 {\cal E}^{\triangleleft}_{\rm Mott}  + o(\tau^4)
\end{equation}
where $E_{\rm Mott}^\diamond =  \frac{U M}{2}n_0 (n_0-1)-\frac{\tau^2}{U}n_0(n_0+1)\,{\rm tr}(t^2)$
is the previously reported result\cite{A:Freericks2}, while ${\cal E}^{\triangleleft}_{\rm Mott} =\frac{n_0}{U^{2}}\, (n_0+1)(2 n_0+1) {\rm tr}(t^3)$
takes into account triangular loops, but it  does not affect the boundary of the Mott lobe, since an identical term appears in the  energy of defect states. 
Indeed, the latter has the form
\begin{equation}
\label{E:def}
E_{\rm def} \!=\! E_{\rm Mott}^\diamond + E_{\rm def}^\diamond + \tau^3\left({\cal E}_{\rm Mott}^{\triangleleft} \!+{\cal E}^{\triangleleft}_{\rm def}\!+{\cal E}^{\rtimes}_{\rm def} \right) + o(\tau^4)
\end{equation}
where
\begin{equation}
E_{\rm def}^\diamond = UC_{\rm def}^{(0)}  - \tau \lambda_1\, C_{\rm def}^{(1)} + \tau^2 {\cal E}_{\rm def}^{(2)}+\tau^3 {\cal E}_{\rm def}^{(3)}
\end{equation}
is the correct third-order result when triangular loops and hopping inhomogeneities are absent.\cite{A:Freericks2} 
In detail, introducing the matrices $\zeta_{j h} \equiv \delta_{j h} \sum_{i=1}^M t_{i j}^2$, $(t^{[p]})_{j h} = t^p_{j h}$ and $ t^{\triangleleft}_{j h}\equiv \left(t^3\right)_{j j} \delta_{j h}$, 
one gets
\begin{equation}
\label{E:Ed2}
{\cal E}_{\rm def}^{(2)} = \frac{ C_{\rm def'}^{(1)} }{2 U} \left[ C_{\rm def}^{(2)} {\mathbf f}_1 \cdot \zeta {\mathbf f}_1 - 2  C_{\rm def}^{(1)}\lambda_1^2 \right],
\end{equation}
 \begin{eqnarray}
\label{E:Ed3}
{\cal E}_{\rm def}^{(3)} &=& \frac{ C_{\rm def}^{(1)} C_{\rm def'}^{(1)}}{U^2} \Big \{ C_{\rm def}^{(3)} \frac{\lambda_1}{4} \,{\mathbf f}_1 \cdot \zeta {\mathbf f}_1  \nonumber\\
&-&(2 n_0+1)\Big [ 2\,{\mathbf f}_1 \cdot t^{[3]} {\mathbf f}_1  + \lambda_1^3 \Big ] \Big \}
\end{eqnarray}
where def' denotes the defect state complementary to def.

As it is clear from table~\ref{T:coeffs}, 
the previously unreported
,\cite{N:triangular}  third-order terms,
\begin{equation}
\label{E:Edl}
{\cal E}^{\triangleleft}_{\rm def} = \frac{C_{\rm def'}^{(1)}C_{\rm def}^\triangleleft}{4 U^2}\, {\mathbf f}_1 \cdot t^{\triangleleft} \,{\mathbf f}_1
\end{equation}
appearing whenever triangular loops are present, and
\begin{equation}
\label{E:Edi}
{\cal E}^{\rtimes}_{\rm def} =  -\frac{\left[C^{(2)}_{\rm def}C^{(1)}_{\rm def'}\right]^2}{4 U^2 C^{(1)}_{\rm def}} \sum_{k=2}^{M} \frac{\left( {\mathbf f}_1 \cdot \zeta {\mathbf f}_k\right)^2}{\lambda_1-\lambda_k}
\end{equation}
arising from topological inhomogeneity, can contribute significantly to the perturbative expansion of $E_{\rm def}$, and hence of Eqs.~(\ref{E:mu}). 
Note that, similar to the previously known  terms,\cite{A:Freericks2}  ${\cal E}^{\triangleleft}_{\rm def}$ depends on the maximal eigenvector of $t$ only. Conversely
${\cal E}^{\rtimes}_{\rm def}$ depends on all of the eigenvectors ${\mathbf f}_k$ and eigenvalues $\lambda_k$ of $t$. Of course ${\cal E}^{\triangleleft}_{\rm def}$ vanishes on structures lacking of triangular loops, where $(t^3)_{j j}=0$. Likewise, ${\cal E}^{\rtimes}_{\rm def}$ vanishes on structures where $\zeta = z {\mathbb I}$, such as regular lattices
,\cite{N:Kagome} since ${\mathbf f}_h \cdot {\mathbf f}_k = \delta_{h k}$.

\begin{table}
\begin{center}
\begin{tabular}{|c||c|c|}
\hline
$x$  & $C_{\rm part}^x$ & $C_{\rm hole}^x$ \\
\hline 
\hline
(0) & $n_0$ & $1-n_0$ \\
\hline
(1) & $(n_0+1)$   & $n_0$  \\
\hline
(2) & $5 n_0+ 4$ & $5 n_0 + 1$ \\
\hline
(3) & $25 n_0+ 14$ & $25 n_0 + 11$ \\
\hline
$\triangleleft$ & $ 12 + 42\, n_0  + 31\, n_0^2$ & $1+20\, n_0 + 31\, n_0^2$ \\
\hline
\end{tabular}
\end{center}
\caption{\label{T:coeffs}Values of the coefficients  in Eqs.~(\ref{E:Ed2})-(\ref{E:Edi}). Their correctness  has been verified on small arrays  ($M\leq10$) by checking  Eq.~(\ref{E:def}) against exact numerical diagonalization. } 
\end{table}

In the following we sketch the key steps involved in carrying out the results in Eqs.~(\ref{E:Mott})-(\ref{E:Edi}). 
As we mention above, the ground state energies appearing in Eqs.~(\ref{E:mu}) 
are obtained restricting the BH Hamiltonian to the
relevant number eigenspace and regarding the hopping term $V=- \sum_{j h} a_j t_{j h} a_h^+$ as a perturbation on the  {\it atomic} unperturbed Hamiltonian, $H_0 = \frac{U}{2} \sum_j n_j (n_j-1)$. According to standard perturbation theory, the ground state of $H=H_0+\tau V$ and the relevant energy are expanded in powers of the perturbative parameter $\tau$, $|\Psi\rangle = \sum_n \tau^n|\psi_n\rangle$, $E = \sum_n \tau^n \epsilon_n$. The perturbative corrections, obeying the hierarchy of equations
\begin{equation}
\label{E:he}
H_0 |\psi_{n}\rangle+V|\psi_{n-1}\rangle = \sum_{k=0}^{n} \epsilon_k |\psi_{n-k}\rangle,
\end{equation}
are then expanded on the eigenfunctions of  $H_0$. Note that the ground 
state of the latter is non-degenerate for commensurate boson 
populations, $N=n_0 M$, and  $M$-fold degenerate in the number 
eigenspaces relevant to defect states, $N=n_0\, M\pm 1$. Hence we conveniently set $|\psi_n \rangle = \sum_{k=1}^D G_n^{(k)} |k \rangle + \sum_\alpha \Gamma_n^{(\alpha)} |\alpha \rangle $, where the quantities denoted by Latin  letters refer to the $D$-dimensional lowest eigenspace of $H_0$, while those denoted by Greek letters refer to the remaining eigenstates. That is $D_{\rm Mott}=1$, $D_{\rm def}=M$ and $H_0 |x\rangle = u_x |x\rangle$, with $u_k = \bar u < u_\alpha$ $\forall \alpha,k$.  Furthermore, for defect states, we set $|k \rangle = [C_{\rm def}^{(1)}]^{-1/2} \sum_{j=1}^M f_{k j}\, b_j^{\rm def} |n_0\rangle\rangle $, where $f_{k j}$ is the $j$-th component of ${\mathbf f}_k$, $b_j^{\rm hole}= a_j$, $b_j^{\rm part}= a_j^+$, and $|n_0\rangle\rangle = \bigotimes_{j=1}^M |n_0\rangle_j$ is the non-degenerate ground state of $H_0$ in the commensurate-filling case, $N=M n_0$. This makes  the first perturbative corrections quite straightforward. Indeed, one gets  $|\psi_0\rangle = |1\rangle$, $\epsilon_0 = \bar u$ and $\epsilon_1^{\rm def} = - C_{\rm def}^{(1)} \lambda_1$, where ``1'' henceforth stands for $k=1$. The subsequent corrections are obtained by projecting Eq.~(\ref{E:he}) on suitably chosen eigenvectors of $H_0$, with the  assumption that $\|\sum_{n=1}^{m} \tau^n |\psi_n\rangle \|^2 = 1+o(\tau^{m+1})$.

Here we focus on the third order correction  for defect states, containing the previously unreported terms in Eqs.~(\ref{E:Edl}) and (\ref{E:Edi}). 
 Projecting   Eq.~(\ref{E:he}) with $n=2$ onto $|\psi_0\rangle = |1\rangle$ one gets $\epsilon_3^{\rm def} = \sum_\alpha \langle 1|V| \alpha\rangle \Gamma_2^{(\alpha)}$, where $\Gamma_2^{(\alpha)} = \langle \alpha |V - \epsilon_1^{\rm def} | \psi_1 \rangle /(\bar u - u_\alpha)$ is in turn obtained setting $n=1$ and projecting onto $| \alpha\rangle$. Finally, projecting Eq.~(\ref{E:he}) with $n=0$ and $n=1$ onto $| \alpha\rangle$ and $| k\rangle$, respectively, gives the coefficients for the first order correction, $\Gamma_1^{(\alpha)} = \langle\alpha|V|1\rangle/(\bar u - u_\alpha) $ and
\begin{equation}
\label{E:G1}
G_1^{(k)}=\frac{\sum_\alpha \langle k |V|\alpha\rangle \Gamma_1^{(\alpha)}}{C_{\rm def}^{(1)}(\lambda_k-\lambda_1)}=\frac{C_{\rm def}^{(2)}C_{\rm def'}^{(1)}}
{2 U C_{\rm def}^{(1)}} \frac{{\mathbf f}_1 \cdot \zeta {\mathbf f}_k}{\lambda_k\! -\!\lambda_1}
\end{equation}
Putting these results together gives
\begin{eqnarray}
\label{E:cor3}
\epsilon_3^{\rm def} &=& \sum_{\alpha \beta} \Gamma_1^{(\alpha)} \langle \alpha|V-\epsilon_1^{\rm def}|\beta\rangle \,\Gamma_1^{(\beta)} \nonumber\\
&+& C_{\rm def}^{(1)}\sum_{k=2}^M (\lambda_k -\lambda_1) \left[G_1^{(k)}\right]^2 
\end{eqnarray}
whose first term provides the third order correction in Eq.~(\ref{E:Edl}), as well as those in Eq.~(\ref{E:Ed3}). The second term in Eq.~(\ref{E:cor3}) is instead responsible for the ``topological'' correction in  Eq.~(\ref{E:Edi}). Note that the latter is the lowest order term where the degeneracy of the ground state of $H_0$ plays an explicit role.  Indeed, the $|k\rangle$'s with $k>1$ do not contribute energy corrections at lower orders. 
In the non-degenerate case, $N=M n_0$, the topological correction is absent also at the third order, the counterpart of Eq.~(\ref{E:cor3}) featuring the first term only.

In the case of a physically significant quantity such as the local density of states,  the ``topological terms'' $G_1^{(k)}$ in Eq.~(\ref{E:G1}) provide a perturbative correction at an order as low as the first. Indeed, a straightforward calculation shows that
\begin{eqnarray}
\rho_j \!\! &=&\! \!  \langle\psi| n_j|\psi\rangle=\langle\psi_0 | n_j|\psi_0 \rangle+2 \tau \langle\psi_0 | n_j|\psi_1\rangle + o(\tau^3) \nonumber\\
\!\!&\approx&\!\!  \langle 1| n_j|1 \rangle+2 \tau \sum_{k=2}^{M} \langle1 | n_j|k\rangle G_1^{(k)}  \nonumber \\
\label{E:rho2}
\!\!&=&\!\! 
n_0+\sigma_{\rm def} f_{1 j}\!\!\left[f_{1 j} \!-\!\tau \frac{C_{\rm def}^{(2)}C_{\rm def'}^{(1)}}{U C_{\rm def}^{(1)}}\sum_{k=2}^M \frac{f_{k j}\,{\mathbf f}_1\!\! \cdot\! \zeta {\mathbf f}_k}{\lambda_1 \!\!-\!\!\lambda_k}\right]
\end{eqnarray}
where $\sigma_{\rm part} =-\sigma_{\rm hole} =  1 $. A comparison with exact numerical diagonalization on small arrays ($M \leq 10$) shows that Eq.~(\ref{E:rho2}) provides satisfactory results. A similar correction appears also in the two-point correlation, $\langle\psi| a_j a_h^+|\psi\rangle$, which we do not discuss here.
\begin{figure}
\begin{center}
\includegraphics[width=8.5cm]{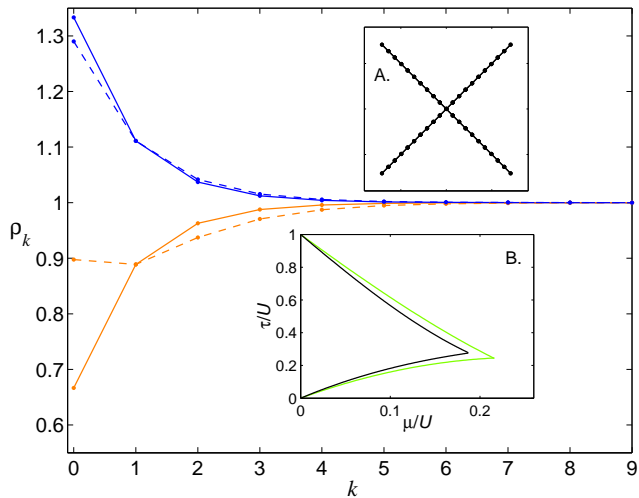}
\caption{\label{F:s4}(color online) {\bf A}: Four fold star network; {\bf B}: first Mott lobe for the four-fold star network (dark) and the 1D lattice (lighter hue, for comparison); {\bf Main plot}: Local density for the ground state of the four-fold star network as a function of the {\it chemical distance} from the central site, $k$. Dark and light hues refer to particle and hole defect states, respectively. Solid and dashed lines refer to the zeroth and first order perturbative results ($\tau/U=0.1$).}
\end{center}
\end{figure}

The first order correction to the local density arising from topological inhomogeneity is clearly shown in Fig.~\ref{F:s4}, where we considered the case of a four-fold star network\cite{A:Burioni,A:Brunelli} (see figure caption for details).

\begin{figure}[t!]
\begin{center}
\includegraphics[width=8.5cm]{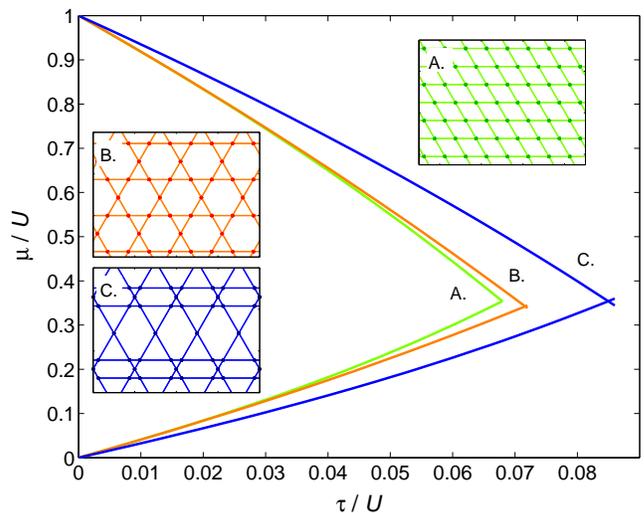}
\caption{\label{F:z4} (color online) Three different 2D structures and, in the same hue, the first Mott lobe thereof. For the inhomogeneous Kagom\`e lattice (C.) we choose the hopping matrix so that the average hopping amplitude is one. In detail $t_{h k}$ is either 0.8 (long links) or 1.2 (short links).}
\end{center}
\end{figure}
 
In Figure \ref{F:z4} we compare the first Mott lobes for three different 2D structures, namely the Euclidean lattice (A.), the homogeneous (B.) and an inhomogeneous (C.) Kagom\`e lattice\cite{A:Santos}. Note that the first two (homogeneous) structures have the same coordination number, $z=4$, so that the only difference between their Mott lobes, Eqs.~(\ref{E:mu}),
 comes from the presence of triangular loops in the latter. The inhomogeneous Kagom\`e lattice has the same coordination as A. and B., but it features two kinds of hopping amplitudes and three kinds of sites (see figure caption for details). Hence its Mott lobes get contributions from both  Eq.~(\ref{E:Edl}) and Eq.~(\ref{E:Edi}).    

\section{Conclusions}
In summary,  we report the complete third order expansion for the ground state of the BH model on a completely arbitrary structure. We show that in the general case two kinds of previously unreported\cite{N:triangular} terms appear. The first kind arises from the presence of triangular loops,
and contributes to the ground state energy of both commensurate-filling and defect states.  Due to exact cancelation only the latter influences the boundaries of the Mott lobes. The second kind of contribution occurs in the defect states of systems characterized by inhomogeneous hopping. As we mention above, these include superlattices with constant coordination or  hopping amplitude, such as those considered in Fig.~\ref{F:z4} (C.) or in Ref.~\onlinecite{A:Cataudella}, respectively. Quite interestingly, hopping inhomogeneity produces perturbative terms that depend on the entire spectrum of the hopping matrix rather than on its maximal eigenpair. Furthermore, other than a third order correction to the ground state energy, it gives rise to a first order correction in the local density of bosons.
We remark that, following the procedure sketched in Ref.~\onlinecite{A:Niemeyer}, our results can be adapted to field-tuned transitions in a superlattice of Josephson junctions.\cite{A:Cataudella}
More in general they supply a basis for a comprehensive study aimed at a deeper understanding of the role of topology in quantum phase transitions\cite{A:BHComb}. 

\acknowledgments
The work of P.B. has been entirely 
supported by MURST project {\it Quantum Information
  and Quantum Computation on Discrete Inhomogeneous Bosonic
  Systems}. A.V. also acknowledges partial financial support from the same
project.


\begin{thebibliography}{10}

\bibitem{A:Fisher}
M.~Fisher, P.~Weichman, G.~Grinstein and D.~S. Fisher, Phys. Rev. B {\bf 40},
  546 (1989).

\bibitem{A:Jaksch}
D.~Jaksch, C.~Bruder, J.~Cirac, C.~Gardiner and P.~Zoller, Phys. Rev. Lett.
  {\bf 81}, 3108 (1998).

\bibitem{A:Greiner}
M.~Greiner, I.~Bloch, O.~Mandel, T.~W. Hansch and T.~Esslinger, Phys. Rev.
  Lett. {\bf 87}, 160405 (2001).

\bibitem{A:Kashurnikov02}
V.~Kashurnikov, N.~Prokof'ev and B.~Svistunov, Phys. Rev. A {\bf 66}, 031601(R)
  (2002).

\bibitem{A:Batrouni02}
G.~Batrouni, V.~Rousseau, R.~Scalettar, M.~Rigol, A.~Muramatsu, P.~J.~H.
  Denteneer and M.~Troyer, Phys. Rev. Lett. {\bf 89}, 117203 (2002).

\bibitem{A:Rey}
A.~Rey, K.~Burnett, R.~Roth, M.~Edwards, C.~Williams and C.~Clark, J. Phys. B
  {\bf 36}, 825--841 (2003).

\bibitem{A:Pupillo03}
G.~Pupillo, E.~Tiesinga and C.~J. Williams, Phys. Rev. A {\bf 68}, 063604
  (2003).

\bibitem{CM:Jain}
P.~Jain and C.~Gardiner, J. Phys. B {\bf 37}, 3649 (2004).

\bibitem{A:Guidoni97}
L.~Guidoni, C.~Trich{\'{e}}, P.~Verkerk and G.~Grynberg, Phys. Rev. Lett. {\bf
  79}, 3363--3366 (1997).

\bibitem{A:Roth03}
R.~Roth and K.~Burnett, Phys. Rev. A {\bf 68}, 023604 (2003).

\bibitem{A:Peil}
S.~Peil, J.~V. Porto, B.~L. Tolra, J.~M. Obrecht, B.~E. King, M.~Subbotin,
  S.~L. Rolston and W.~D. Phillips, Phys. Rev. A {\bf 67}, 051603(R) (2003).

\bibitem{A:Blakie}
P.~B. Blakie and C.~W. Clark, J. Phys. B {\bf 37}, 1391 (2004).

\bibitem{A:Santos}
L.~Santos, M.~Baranov, J.~Cirac, H.-U. Everts, H.~Fehrmann and M.~Lewenstein,
  Phys. Rev. Lett. {\bf 93}, 030601 (2004).

\bibitem{A:Grier}
D.~Grier, Nature {\bf 424}, 810 (2003).

\bibitem{A:Folman01}
R.~Folman and J.~Schmiedmayer, Nature {\bf 413}, 466 (2001).

\bibitem{A:FazioPR}
R.~Fazio and H.~van~der Zant, Phys. Rep. {\bf 355}, 235 (2001).

\bibitem{A:Meyer02}
R.~Meyer, S.~E. Korshunov, C.~Leemann and P.~Martinoli, Phys. Rev. B {\bf 66},
  104503 (2002).

\bibitem{A:Cataudella}
V.~Cataudella and R.~Fazio, Europhys. Lett. {\bf 61}, 341 (2003).

\bibitem{A:BHComb}
P.~Buonsante, R.~Burioni, D.~Cassi, V.~Penna and A.~Vezzani, 
cond-mat/0405520, Phys. Rev. B {\bf 70}, 22XXXX (to be published 1 
December 2004).

\bibitem{A:CombL}
R.~Burioni, D.~Cassi, I.~Meccoli, M.~Rasetti, S.~Regina, P.~Sodano and
  A.~Vezzani, Europhys. Lett. {\bf 52}, 251 (2000).

\bibitem{A:MesoComb}
P.~Buonsante, R.~Burioni, D.~Cassi and A.~Vezzani, Phys. Rev. B {\bf 66},
  094207 (2002).

\bibitem{A:Sheshadri}
K.~Sheshadri, H.~Krishnamurthy, R.~Pandit and T.~Ramakrishnan, Europhys. Lett.
  {\bf 22}, 257 (1993).

\bibitem{A:Amico}
L.~Amico and V.~Penna, Phys. Rev. Lett. {\bf 80}, 2189 (1998).

\bibitem{A:vanOosten}
D.~van Oosten, P.~van~der Straten and H.~Stoof, Phys. Rev. A {\bf 63}, 053601
  (2001).

\bibitem{A:LobiMF}
P.~Buonsante and A.~Vezzani, Phys. Rev. A {\bf 70}, 033608 (2004).

\bibitem{A:Kuehner}
T.~K{\"{u}}hner and H.~Monien, Phys. Rev. B {\bf 58}, R14741 (1998).

\bibitem{A:Elstner99a}
N.~Elstner and H.~Monien, Phys. Rev. B {\bf 59}, 12184 (1999).

\bibitem{A:Batrouni}
G.~Batrouni, R.~Scalettar and G.~Zimanyi, Phys. Rev. Lett. {\bf 65}, 1765
  (1990).

\bibitem{A:Kashurnikov96}
V.~Kashurnikov, A.~Krasavin and B.~Svistunov, JETP Lett. {\bf 64}, 99 (1996).

\bibitem{A:Batrouni92}
G.~Batrouni and R.~Scalettar, Phys. Rev. B {\bf 46}, 9051--9062 (1992).

\bibitem{A:Batrouni93}
G.~Batrouni, B.~Larson, R.~Scalettar, J.~Tobochnik and J.~Wang, Phys. Rev. B
  {\bf 48}, 9628--9635 (1993).

\bibitem{A:Kisker97}
J.~Kisker and H.~Reiger, Phys. Rev. B {\bf 55}, R11981 (1997).

\bibitem{A:Lee01}
J.-W. Lee, M.-C. Cha and D.~Kim, Phys. Rev. Lett. {\bf 87}, 247006 (2001).

\bibitem{A:Freericks2}
J.~K. Freericks and M.~Monien, Phys. Rev. B {\bf 53}, 2691--2700 (1996).

\bibitem{A:Niemeyer}
M.~Niemeyer, J.~K. Freericks and M.~Monien, Phys. Rev. B {\bf 60}, 2357 (1999).

\bibitem{N:triangular}
A term quite similar to Eq.~(\ref{E:Edl}) is reported in
  Ref.~\onlinecite{A:Niemeyer} addressing the particular case of the regular
  triangular lattice, albeit in the presence of a magnetic field.

\bibitem{N:Kagome}
Note that this can happen also on structures with inhomogeneous hopping, such
  as the {\it trimerized} Kagom\`e lattice\cite{A:Santos}.

\bibitem{A:Burioni}
R.~Burioni, D.~Cassi, M.~Rasetti, P.~Sodano and A.~Vezzani, J. Phys. B {\bf
  34}, 4697 (2001).

\bibitem{A:Brunelli}
I.~Brunelli, G.~Giusiano, F.~P. Mancini, P.~Sodano and A.~Trombettoni, J. Phys.
  B {\bf 37}, S275 (2004).

\end{thebibliography}
\end{document}